\documentclass{article}

     \usepackage[preprint]{neurips_2019}

\usepackage{rotating}
\usepackage[utf8]{inputenc} 
\usepackage[T1]{fontenc}    
\usepackage{hyperref}       
\usepackage{url}            
\usepackage{booktabs}       
\usepackage{amsfonts}       
\usepackage{nicefrac}       
\usepackage{microtype}      
\usepackage{graphicx}
\usepackage{amsmath}
\usepackage{floatrow}
\newfloatcommand{capbtabbox}{table}[][\FBwidth]

\title{ODE-based Deep Network for MRI Reconstruction}
\author{Ali Pour Yazdanpanah\textsuperscript{1,2} \ \ Onur Afacan\textsuperscript{1,2} \ \ Simon K. Warfield\textsuperscript{1,2}\\
\textsuperscript{1}Harvard Medical School, Boston, MA.\\
\textsuperscript{2}Computational Radiology Laboratory, Boston Children's Hospital, Boston, MA.\\
{\tt\small \{ali.pouryazdanpanahkermani,onur.afacan,simon.warfield\}@childrens.harvard.edu}
}

\begin{document}

\maketitle
\begin{abstract}
Fast data acquisition in Magnetic Resonance Imaging (MRI) is vastly in demand and scan time directly depends on the number of acquired k-space samples. The data-driven methods based on deep neural networks have resulted in promising improvements, compared to the conventional methods, in image reconstruction algorithms. The connection between deep neural network and Ordinary Differential Equation (ODE) has been observed and studied recently. The studies show that different residual networks can be interpreted as Euler discretization of an ODE. In this paper, we propose an ODE-based deep network for MRI reconstruction to enable the rapid acquisition of MR images with improved image quality. Our results with undersampled data demonstrate that our method can deliver higher quality images in comparison to the reconstruction methods based on the standard UNet network and Residual network.
\end{abstract}
\section{Introduction}
The connection between deep neural network and the ordinary differential equation has been studied and discussed in different recent works [1,2,3,4,5,6]. It has been shown that residual networks such as ResNet [7] and recurrent neural network decoders can be modeled as a discretization of a continuous ODE model. An ODE-based model and its relation to the residual network can be shown as follows:
\begin{equation} \label{Eq1}
\begin{aligned}
 \ \ \text L_{t_1} =\text L_{t_0} + f(\text L_{t_0},\theta) \ \ \ \  \ \ \ \ \ \ \ \  \ \ \ \ \ \ \ \ \ \ \ \ (ResNet)
\end{aligned}
\end{equation}
\begin{equation} \label{Eq2}
\begin{aligned}
\text L_{t_1} =\text L_{t_0} + \int_{t_0}^{t_1} f(\text L_{t},\theta)\ dt \ \ \ \  \ \ \ \ \ \ \ \ (ODE)
\end{aligned}
\end{equation}
where $\text L_{t_0}, \text L_{t_1}$ are the residual block input and output. $f$ represents the network-defined nonlinear operator which preserves the dimensionality of $\text L_{t_0}$ and $\theta$ represents the network weights. The defined ODE ($\frac{d\text L}{dt} = f(\text L_{t},\theta)$) is described in terms of its solution in $t=t_1$. 
The forward step of ODE Euler discretization is as follows:
\begin{equation} \label{Eq3}
\begin{aligned}
 \ \ \text L_{t_{0}+h} =\text L_{t_0} + hf(\text L_{t_0},\theta) \ \ \ \  \ \ \ \ \ \ \ \  \ \ \ \ \ \ \ \ \ \ \ \ \ \ \ \ \ \ \ \ \ \ \ \ \ \ \
\end{aligned}
\end{equation}
It can be observed that the single forward step of Eq~\ref{Eq3}, can be considered as the equivalent to the formulation of the residual block. Therefore, The ODE discretization model can lead to different ODE-inspired network architectures. In this paper, we present an ODE-based deep network for MRI reconstruction which extends the conventional reconstruction framework to its data-adaptive variant using the ODE-based network and provides an end-to-end reconstruction scheme. We evaluate the reconstruction performance of our method to the reconstruction methods based on the standard UNet network [8] and Residual network. 
\section{Method}
The discretized version of MR imaging model given by		
\begin{equation} \label{Eq4}
\begin{aligned}
\text d=\text E \text x + \text n.
\end{aligned}
\end{equation}
where $\text x$ is the samples of unknown MR image, and $\text d$ is the undersampled k-space data. $\text E =\text {FS}$ is an encoding matrix, and $\text F$ is an undersampled Fourier operator. $\text S$ is a matrix representing the sensitivity map of the coils, and $\text n$ is noise. Assuming that the interchannel noise covariance has been whitened, the reconstruction relies on the least-square approach:
\begin{equation} \label{Eq5}
\begin{aligned}
\hat{\text x} =\underset{\text x}{ argmin} \ \|\text d-\text E\text x\|_{2}^{2}
\end{aligned}
\end{equation}
The ODE-based reconstruction framework we used for solving the Eq~\ref{Eq5} is shown in Fig~\ref{fig1}. \\
For a conventional neural network, we minimize the loss function ($l$) over a set of training pairs and we search for the weights ($\theta$) that minimize that loss function:
\begin{equation} \label{Eq6}
\begin{aligned}
\underset{\theta}{ minimize} \ \frac{1}{M} \sum\limits_{i=1}^M l(L(\theta;x_i,y_i)) + \text R(\theta)
\end{aligned}
\end{equation}
where $(x_i,y_i)$ is the $i_{th}$ training pairs (input and ground truth). $R$ is a regularization operator and $M$ is the number of training pairs. The loss function depends implicitly on $\theta$. This optimization is usually solved through Stochastic Gradient Descent (SGD) and backpropagation to compute the gradient of $L$ with respect to $\theta$. In our ODE-based network, besides the $L$s, the network weights also change with respect to time as well. In this case, we need to solve the following constrained optimization problem :
\begin{equation} \label{Eq7}
\begin{aligned}
\underset{p,\theta}{ minimize} \ \frac{1}{M} \sum\limits_{i=1}^M l(L_{t_1};x_i,y_i) + \text R(p,\theta)
\end{aligned}
\end{equation}
\begin{equation} \label{Eq8}
\begin{aligned}
\frac{d\text L}{dt} = f(\text L_{t},\theta_{t})
\end{aligned}
\end{equation}
\begin{equation} \label{Eq9}
\begin{aligned}
\frac{d\theta}{dt} = w(\theta_{t},p)
\end{aligned}
\end{equation}
where $\theta_t$ is parameterized by the learnable dynamics of
\begin{equation} \label{Eq10}
\begin{aligned}
\theta_{t_1} =\theta_{t_0} + \int_{t_0}^{t_1} w(\theta_{t},p)\ dt
\end{aligned}
\end{equation}
where $w$ is a nonlinear operator responsible for the network weights dynamics and $p$ is the parameters for $w$.
We also augment the learnable space and solve the ODE flow as follows so that the learned ODE representation won't preserve the input space topology [5].
\begin{equation} \label{Eq11}
\begin{aligned}
\frac{d}{dt} \begin{bmatrix}
    \text L\\
    a
\end{bmatrix} = f(\begin{bmatrix}
    \text L_{t}\\
    a_{t}
\end{bmatrix},\theta_{t})
\end{aligned}
\end{equation}
where $a_{0}=0$. We use the discretize-then-optimize method [4,6] to calculate the gradients for backpropagating through ODE layers. Figure \ref{fig2} shows the proposed ODE-based deep network. Five residual blocks have been used in our method (N=5).
\section{Results and Discussion}
\begin{figure}
\begin{floatrow}
\ffigbox{%
  \includegraphics[scale=0.37]{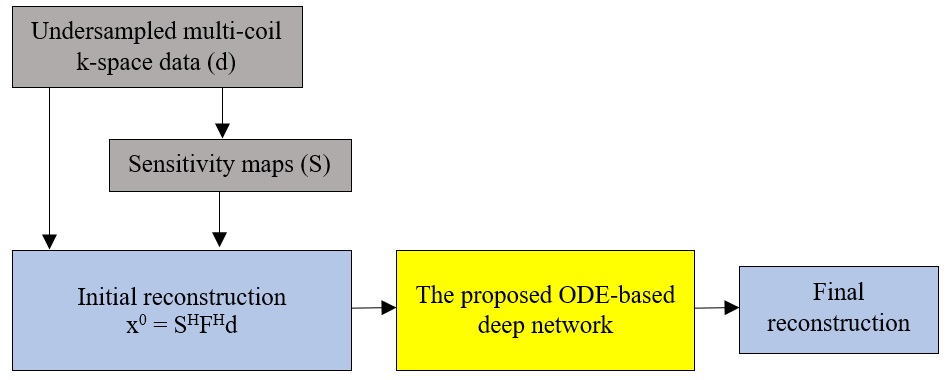}%
  }{%
  \caption{The reconstruction framework}%
  \label{fig1}}
\capbtabbox{%
\begin{tabular}{rllll}
\hline
\multicolumn{3}{c} {Brain Dataset} \\
\cline{2-3} 
\cline{4-5} 
Method & PSNR  & SSIM   \\
\hline
Proposed & $54.5\pm1.37$  & $0.99\pm0.0063$   \\
UNet & $52.4\pm1.54$  & $0.98\pm0.0075$   \\
ResidualNet & $50.1\pm1.65$  & $0.978\pm0.0097$   \\
\hline
\end{tabular}
}
{
\caption{PSNR and SSIM variations on Brain dataset}%
\label{table1}
}
\end{floatrow}
\end{figure}

\begin{figure}
  \centering
  \centerline{\includegraphics[scale=0.5]{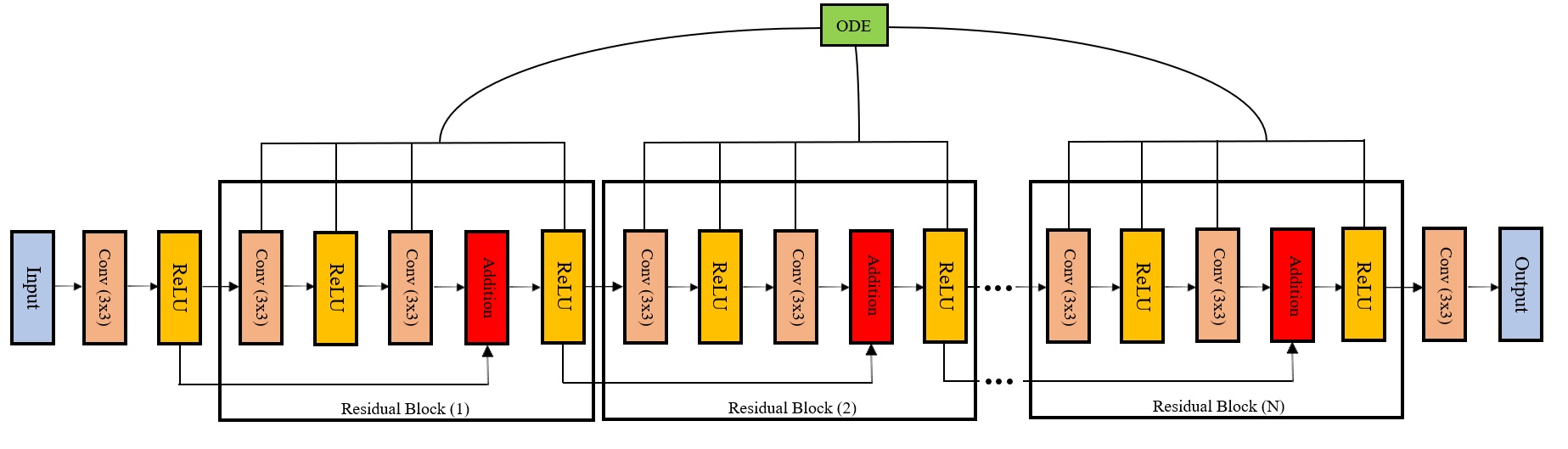}}
  \caption{The proposed ODE-based deep network.}
  \label{fig2}
\end{figure}

\begin{figure}
  \centering
  \centerline{\includegraphics[scale=0.15]{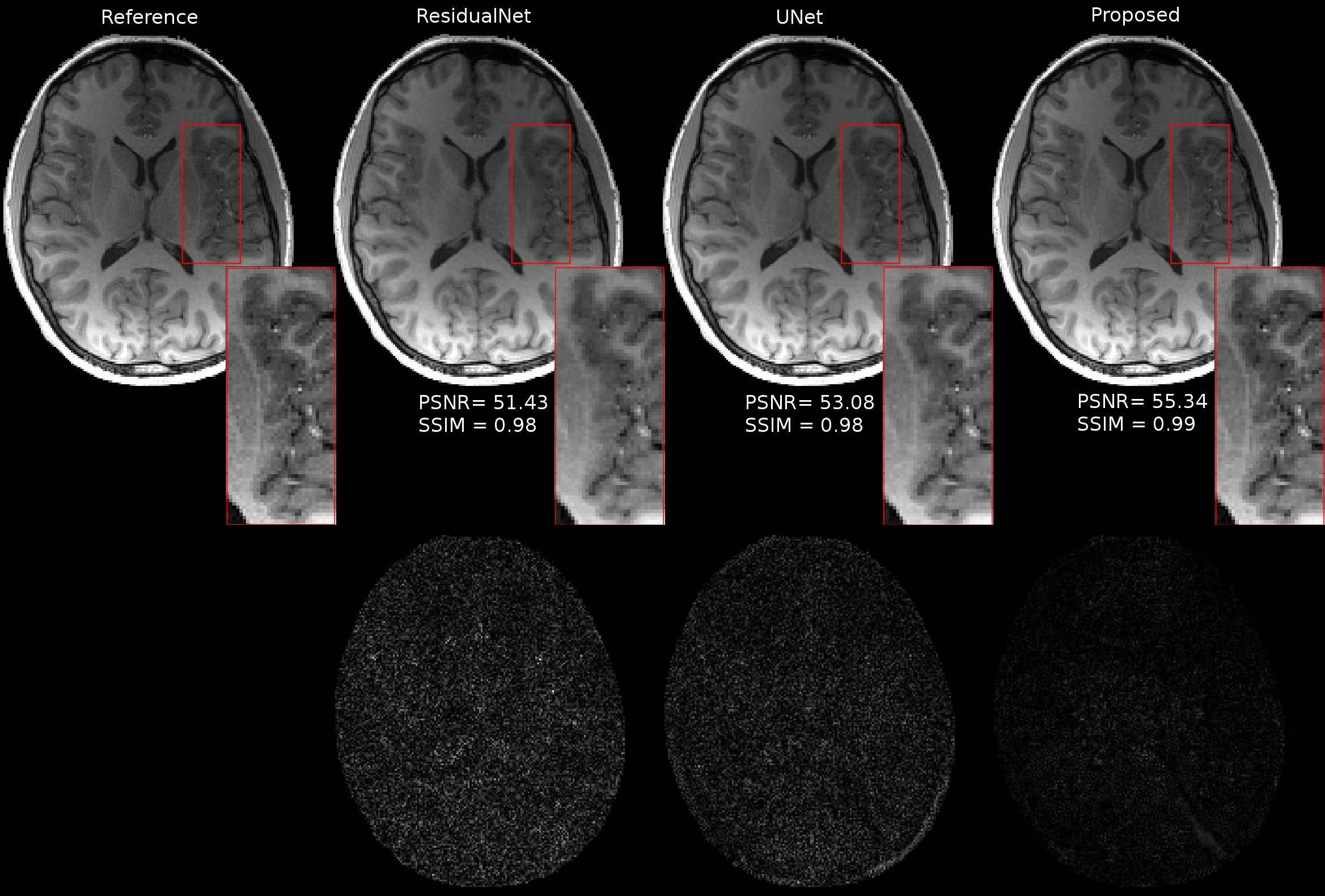}}
  \caption{First row (left to right): Reference image using fully sampled data, ResidualNet reconstruction, UNet reconstruction, and our reconstruction all with undersampling factor of 2x2. Second row includes error maps correspond to each reconstruction results for comparison.}
  \label{fig3}
\end{figure}

In our experiments, we have tested our method with our MPRAGE brain datasets. The data on ten volunteers with a total of 750 brain images used as the training set. Images from fifteen different volunteers have used as the testing set. The sensitivity maps were computed from a block of size 24x24 using the ESPIRiT [9] method. Reconstruction results with the undersampling factor of 2x2 for different approaches are shown in Fig \ref{fig3}. ResidualNet includes the same number of residual blocks as our proposed method (without ODE layers).
Table \ref{table1} shows
that our method consistently has higher Peak Signal-to-noise Ratios (PSNR) and structural similarities (SSIM) compared to the reconstructions using the other two networks. In conclusion, an ODE-based deep network for MRI reconstruction is proposed. It enables the rapid acquisition of MR images with improved image quality. The proposed ODE-based network can be easily adopted by unrolled optimization schemes for better MRI reconstruction accuracy.

\newpage

\subsubsection*{Acknowledgments}

This research was supported in part by NIH grants R01 NS079788, R01 EB019483, R01 DK100404, R44 MH086984, IDDRC U54 HD090255, and by a research grant from the Boston Children's Hospital Translational Research Program.

\section*{References}

[1] Haber, E. and Ruthotto, L., “Stable architectures for deep neural networks,” Inverse Problems, 34(1), 014004 (2017).

[2] Ruthotto,  L.  and  Haber,  E.,  “Deep  neural  networks  motivated  by  partial  differential  equations,” arXiv preprint arXiv:1804.04272 (2018).

[3] Chen, T. Q., Rubanova, Y., Bettencourt, J., and Duvenaud, D. K., “Neural ordinary differential equations,” in [Advances in neural information processing systems], 6571–6583 (2018).

[4] Gholami, A., Keutzer, K., and Biros, G., “Anode:  Unconditionally accurate memory-efficient gradients for neural odes,” arXiv preprint arXiv:1902.10298 (2019).

[5] Dupont,  E.,  Doucet,  A.,  and  Teh,  Y.  W.,  “Augmented  neural  odes,” arXiv  preprint  arXiv:1904.01681 (2019).

[6] Zhang,  T.,  Yao,  Z.,  Gholami,  A.,  Keutzer,  K.,  Gonzalez,  J.,  Biros,  G.,  and Mahoney,  M.,  “Anodev2:  A coupled neural ode evolution framework,” arXiv preprint arXiv:1906.04596 (2019).

[7] He, K., Zhang, X., Ren, S., and Sun, J., “Deep residual learning for image recognition,” in [Proceedings of the IEEE conference on computer vision and pattern recognition], 770–778 (2016).

[8] Ronneberger,  O.,  Fischer,  P.,  and  Brox,  T., “U-net:  Convolutional  networks  for  biomedical  image  segmentation,” in [International Conference on Medical image computing and computer-assisted intervention], 234–241, Springer (2015).

[9] M. Uecker, P. Lai, M. J. Murphy, P. Virtue, M. Elad, J. M.Pauly, S. S. Vasanawala, and M. Lustig, “Espirit an eigenvalue approach to autocalibrating parallel mri:  where sense meets grappa,” Magnetic resonance in medicine, 71(3), 990–1001 (2014).

\end{document}